# Lipidation-induced bacterial cell membrane translocation of star-peptides

Amal Jayawardena[1], Andrew Hung[2], Greg Qiao[3], Elnaz Hajizadeh[1*]

[1] Soft Matter Informatics Research Group, Department of Mechanical Engineering, Faculty of Engineering and Information Technology, University of Melbourne, Parkville, VIC, 3010, Australia
[2] School of Science, STEM College, RMIT University, VIC, 3001, Australia
[3] Department of Chemical Engineering, Faculty of Engineering and Information Technology, University of Melbourne, Parkville, VIC, 3010, Australia
*Corresponding Author: ellie.hajizadeh@unimelb.edu.au

## Abstract

The rapid emergence of multidrug-resistant (MDR) bacteria demands development of novel and effective antimicrobial agents. Structurally Nanoengineered Antimicrobial Peptide Polymers (SNAPPs), characterized by their unique star-shaped architecture and potent multivalent interactions, represent a promising solution. This study leverages molecular dynamics simulations to investigate the impact of lipidation on SNAPPs' structural stability, membrane interactions, and antibacterial efficacy. We show that lipidation with hexanoic acid (C6), lauric acid (C12), and stearic acid (C18) enhances the α-helical stability of SNAPP arms, facilitating deeper insertion into the hydrophobic core of bacterial membranes. Among the variants, C12-SNAPP exhibits the most significant bilayer disruption, followed by C6-SNAPP, whereas the excessive hydrophobicity of C18-SNAPP leads to pronounced arm back-folding towards the core, reducing its effective interaction with the bilayer and limiting its bactericidal performance. Additionally, potential of mean force (PMF) analysis reveals that lipidation reduces the free energy barrier for translocation through the bilipid membrane compared to non-lipidated SNAPPs. These findings underscore the critical role of lipidation in optimizing SNAPPs for combating MDR pathogens. By fine-tuning lipid chain lengths, this study provides a framework for designing next-generation antimicrobial agents to address the global antibiotic resistance crisis, advancing modern therapeutic strategies.

## 1. Introduction

The rapid rise of multidrug-resistant (MDR) infections has become a global health crisis, posing a significant threat to public health and modern medicine. The World Health Organization (WHO) has declared antimicrobial resistance (AMR) as a critical priority, predicting that by 2050, AMR could lead to 10 million deaths annually if no effective solutions are implemented[1-6]. The lack of effective antibiotics in the development pipeline has further exacerbated the issue, necessitating urgent efforts to discover novel antimicrobial agents that can combat both Gram-positive and Gram-negative MDR bacterial strains[7-10].

Structurally Nano Engineered Antimicrobial Peptide Polymers (SNAPPs) represent a transformative breakthrough in addressing the global challenge of multidrug-resistant (MDR) bacteria[11, 12]. Developed by researchers at the University of Melbourne, SNAPPs are star-shaped polymers characterized by multiple arms made up of co-peptides with varying length and amino acid sequences. One of the most effective SNAPPs studied to date is composed of arms that feature an alternating sequence of lysine and valine residues in a 2:1 ratio. Cryo-transmission electron microscopy (cryo-TEM) and extensive laboratory experiments have demonstrated the exceptional efficacy of SNAPPs in targeting and eradicating both Gram-positive and Gram-negative bacteria, including MDR strains[11, 12]. Additionally, our previous molecular dynamics simulations provided valuable insights into the SNAPPs' mechanism of action, which involves the disruption of bacterial membranes and the formation of pores[13].

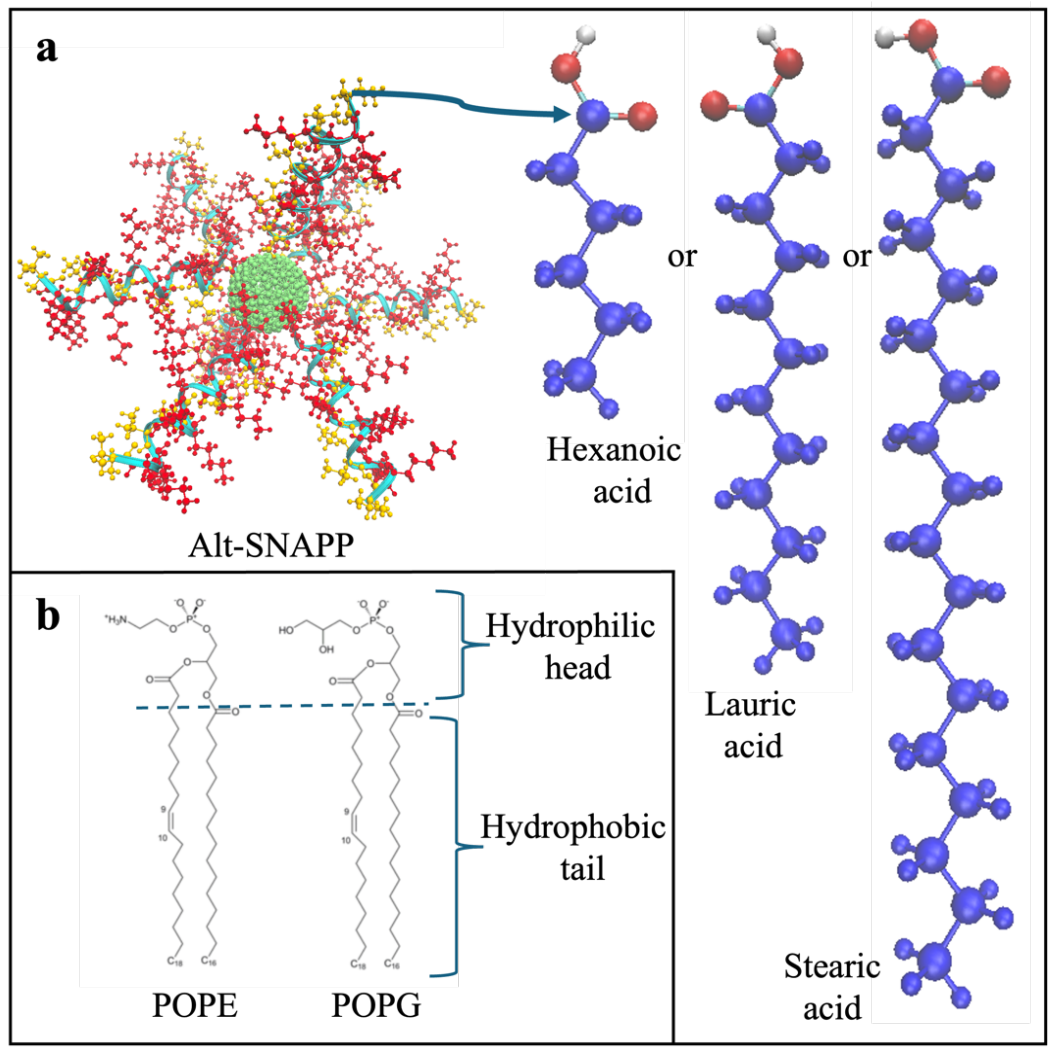

Figure 1 - *An overview of the molecular components and topology of the simulated systems. Lysine residues are shown in red, valine residues in yellow, and the attached lipids (C6, C12, and C18) in blue. (a) Depicts the Corey-Pauling-Koltun (CPK) visualization of the simulated SNAPP models using VMD (Visual Molecular Dynamics). Hexanoic acid's C-terminus is connected to the N-terminus of the terminal valine in the alt-SNAPP model to create C6-SNAPP. Similarly, the C-terminus of lauric acid connects to the N-terminus of valine in alt-SNAPP to form the C12-SNAPP model, while the C-terminus of stearic acid connects to the N-terminus of valine in alt-SNAPP to create the C18-SNAPP model. (b) Chemical structure illustrating the hydrophilic head groups and hydrophobic tails of 1-palmitoyl-2-oleoyl-sn-glycero-3-phosphoethanolamine (POPE) and 1-palmitoyl-2-oleoyl-sn-glycero-3-phosphoglycerol (POPG) lipids used in the bilayer composition.*

While pore formation in bacterial membranes is recognized as a primary mechanism of action for many antimicrobial peptides (AMPs), it is far from the only pathway by which bacterial killing can occur[14-17]. In fact, bacterial eradication does not necessarily rely on the formation of discrete pores in the lipid bilayer[18-23]. Instead, alternative mechanisms such as lipid modification-induced membrane destabilization have been demonstrated to play a significant role in bacterial death[24]. Lipid modifications, such as the conjugation of fatty acids, enhance the hydrophobic interactions between AMPs and bacterial membranes, allowing the peptides to insert deeply into the hydrophobic core of the bilayer[25]. This insertion induces significant membrane stress, destabilizing the bilayer structure and creating mechanical tension that disrupts lipid packing. Such destabilization has been shown to impair membrane-associated processes, such as ion transport and ATP synthesis, leading to energy depletion and cell death[24]. Moreover, this destabilization has been linked to the activation of programmed cell death mechanisms, such as autolysis and the production of reactive oxygen species (ROS), further contributing to bacterial killing[26]. The ability of lipid-modified AMPs to induce such effects without relying on pore formation expands the range of their antibacterial activity, especially against pathogens with adaptive resistance to pore-dependent mechanisms.

In this study, we investigate the impact of lipidating the arms of SNAPP on their structural properties and membrane interaction using molecular dynamics simulations. Specifically, we examine how the attachment of lipid molecules to the termini of SNAPP arms influences their mechanism of action and affects their membrane disruption behaviour. Additionally, we explore how lipidation alters the energy barrier for adsorption and translocation through the bilayer, providing insights into its potential role in modulating SNAPP membrane interactions compared to previously studied SNAPPs[13]. The lipid modifications studied include C6-SNAPP, C12-SNAPP and C18-SNAPP variants, where hexanoic acid, lauric acid and capric acid are attached to the N-terminus of the end valine of each arm of alt-SNAPP respectively as shown in Figure 1a. The alt-SNAPP variant consists of an alternating amino acid sequence of lysine and valine residues (KKVKKVKKVKKVKKV) as shown in Figure 2a[13].

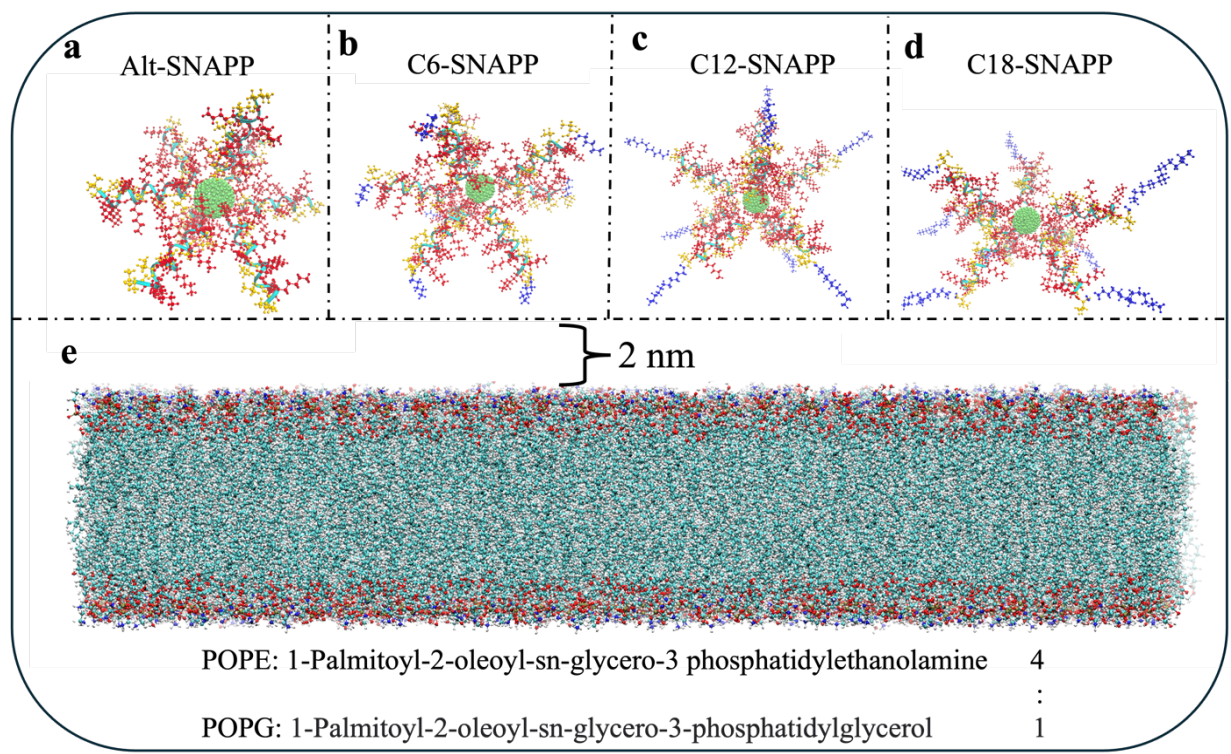

*Figure 2 - The finalized Corey-Pauling-Koltun (CPK) visualization of the 8-arm simulated SNAPP models. Lysine residues are shown in red, valine residues in yellow, and the attached lipids (C6, C12, and C18) in blue. (a) Depicts the simulated system of alt-SNAPP. (b) Shows the C6-SNAPP model, where the attached C6 group is coloured in blue. (c) Illustrates the C12-SNAPP model, with attached lipids coloured in blue. (d) Represents the C18-SNAPP model, with attached lipids also shown in blue. (e) Visualizes the bilayer composed of 1-palmitoyl-2-oleoyl-sn-glycero-3-phosphoethanolamine (POPE) and 1-palmitoyl-2-oleoyl-sn-glycero-3-phosphoglycerol (POPG) in a 4:1 ratio. All SNAPP models were placed with their termini positioned 2 nm above the bilayer.*

Molecular dynamics simulations are essential for understanding the structural and mechanistic differences introduced by lipidation in SNAPP variants[27, 28]. These simulations allow us to observe interactions at the molecular level, offering insights into how lipid modifications influence SNAPP behaviour in hydrophobic and hydrophilic environments[29-31].

This manuscript is organized as follow. Section 2 describes the Methodology and techniques used for modelling and analysis, including molecular dynamics simulations, secondary structure and umbrella sampling. Section 3 outlines the Results and Discussion and is divided into three subsections: 3.1. examines the secondary structure of lipidated SNAPP models; 3.2. investigates the mechanism of action of lipidated SNAPPs; and 3.3. provides theoretical insights into the potential of mean force for lipidated C6, C12 and C18 SNAPP arms in comparison to the alt-SNAPP arm. Finally, the Conclusion section consolidates the findings and underscores their implications for the development of next-generation antimicrobial agents.

## 2. Methodology

### 2.1. MD Simulations

In this study, we used GROMACS 2021.v3 molecular dynamics simulation software package[32] to perform the simulations detailed in Table A1. The CHARMM36 force field was employed, as it has been widely validated in numerous studies for elucidating membrane properties and peptide-lipid interactions[33]. The modelled SNAPP chemical structures, i.e., alt-block SNAPP, C6-SNAPP, C12-SNAPP, and C18-SNAPP all feature eight arms with an alternating sequence of lysine and valine residues in a 2:1 ratio as shown in Figure 2a, 2b, 2c, and 2d. In the lipidated SNAPP variants, lipid molecules are attached to the termini of the valine residues at the end of each arm as shown in Figure 1a.

To construct the structure of an eight-arm SNAPP, a stepwise approach was implemented. Initially, the individual arms of SNAPP were modelled using the Avogadro molecular modelling software, which facilitated the definition of specific amino acid sequences and the initial secondary structure of each arm[34]. Since the study involves atomistic simulations, the arms were initialized with an α-helical secondary structure to observe their environment-dependent conformational changes during simulations. Subsequently, the nanoengineered core of the SNAPP was generated using the Packmol software[35]. The core was approximated as a hydrophobic sphere, composed of 400 carbon atoms to mimic a nonpolar spherical surface. The individual arms were then covalently bonded to the core via the C-terminal carbonyl carbon, which was connected to a designated core atom using a covalent bond with a bond length of 2.5 Å. This step required precise alignment and connectivity to maintain the intended spatial arrangement of the arms around the central core, yielding the final eight-arm alt-SNAPP structure.

To create lipidated SNAPP models (C6, C12, and C18), the initial lipid structures were generated using the Avogadro molecular modelling software. The lipid chains were aligned and positioned using VMD software[36], ensuring that they were covalently attached to the N-terminus of the terminal valine residue in each arm. This process maintained structural consistency across the lipidated variants. To analyse the bilayer thickness and upper leaflet deformations, VMD's membrane plugin[37] was used.

In the laboratory, SNAPPs are synthesized via ring-opening polymerization of lysine and valine N-carboxyanhydrides (NCAs), initiated from the terminal amines of poly(amidoamine) (PAMAM) dendrimers with $NH_2$ functional groups. This synthesis method ensures that the C-terminus of lysine residues is covalently linked to the core, creating the unique star-shaped structure of SNAPPs[11, 12, 38].

The atomistic model of bilipid membrane was constructed using the CHARMM-GUI input generator, specifically employing the membrane builder tool available within this server[39]. To replicate the inner membrane of Gram-negative bacteria, a bilayer was composed of a mixture of 1-palmitoyl-2-oleoyl-sn-glycero-3-phosphatidylethanolamine (POPE) and 1-palmitoyl-2-oleoyl-sn-glycero-3-phosphatidylglycerol (POPG) lipids in a 4:1 ratio as shown in Figure 1b and Figure 2e. This composition was applied symmetrically to both the upper and lower leaflets of the bilayer to accurately mimic the native membrane environment.

To position the SNAPP above the bilipid layer, the Visual Molecular Dynamics (VMD) software was utilized. This allowed for the precise integration of the SNAPP coordinates with the bilipid layer, ensuring that the bottom of the SNAPP structure was placed at a distance of 2 nm from the top of the bilayer as shown in Figure 2e. Maintaining this initial distance minimizes potential bias in the binding orientation and allows the SNAPP structure to reorient naturally at the start of the simulation, consistent with established protocols in the literature[40, 41]. Detailed descriptions of each simulation are provided in Table A1.

Energy minimization of the system was performed using the steepest descent algorithm until the maximum force was less than 1000 kJ·mol$^{-1}$·nm$^{-1}$. Following this step, the pre-equilibrated system underwent additional equilibration for 50 ns. The LINear Constraint Solver (LINCS)[42] algorithm was used to constrain all hydrogen bonds, while the Particle Mesh Ewald (PME)[43] method was employed for electrostatic calculations. During the NPT equilibration steps, the temperature was maintained at 303.15 K using the Berendsen thermostat[44] with a time constant of 1 ps, and the pressure was held at 1 bar using the Berendsen barostat with semi-isotropic pressure coupling, a time constant of 5 ps, and an isothermal compressibility of $4.5 \times 10^{-5}$ (kJ·mol$^{-1}$·nm$^{-3}$)$^{-1}$. During the production runs, the temperature was maintained at 303.15 K using the Nosé-Hoover thermostat[45] with a time constant of 1 ps, and the pressure was kept at 1 bar using the semi-isotropic Parrinello-Rahman barostat[46] with a time constant of 5 ps and an isothermal compressibility of $4.5 \times 10^{-5}$ (kJ·mol$^{-1}$·nm$^{-3}$)$^{-1}$.

## 2.2. Secondary Structure Analysis

Subsequent analyses included secondary structure determination. The ellipticity at 222 nm was calculated using the method described by Hirst and Brooks[47], implemented within the GROMACS helix command. Ellipticity at 222 nm serves as a quantitative measure of α-helical content in proteins, derived from the differential absorption of circularly polarized light at this specific wavelength. A reduction in helical content correlates directly with a loss of ellipticity at 222 nm.

## 2.3. Umbrella Sampling

Free energy profiles for permeation of single peptide arms derived from SNAPP were determined by calculating the potentials of mean force (PMFs) along a single collective variable, defined as the reaction coordinate, which represents the peptide's position relative to the membrane. These profiles were obtained using the umbrella sampling method[48]. Details of the simulated system are provided in Table A1. The alt-SNAPP, C6 SNAPP, C12-SNAPP and C18-SNAPP arms were pulled from its centre of mass through the membrane along the reaction coordinate as shown in Figure 9a, 9b and 9c. For the free energy calculations, 50 windows were employed, with the force constant of the umbrella potential set to 1000 kJ·mol$^{-1}$·nm$^{-2}$ and pull rate of 0.0025 nm/ps. Each window underwent an equilibration phase of 200 ns, followed by a sampling period of 200 ns. The free energy landscapes were constructed using the weighted histogram analysis method (WHAM)[49], providing detailed insights into the energy barriers associated with peptide translocation across the membrane.

## 3. Results and Discussion:

In previous laboratory circular dichroism experiments[50], it was established that SNAPP arms adopt a random coil conformation in hydrophilic environments and an α-helical conformation in hydrophobic environments. In our earlier publication, we demonstrated that the modelled alt-SNAPP exhibits similar behaviour, validating the accuracy and reliability of our computational model[13].

### 3.1. Secondary Structure Analyses

#### 3.1.1. Secondary structure analysis of C6-SNAPP, C12-SNAPP and C18-SNAPP in water and water + TFE

The impact of lipid modifications, i.e., C6, C12, and C18 lipid attachments on the secondary structure of SNAPP molecules in water is shown in Figure 3a. The results demonstrate that all lipidated SNAPP structures exhibit increased α-helicity compared to their non-lipidated counterpart in water. Despite this enhancement, a gradual decline in α-helicity was still observed, suggesting that while lipidated SNAPP arms form a more stable α-helical structure in water, the transition towards a random coil conformation occurs at a slower rate compared to the non-lipidated arms.

Figure 3b shows that the lipidated SNAPP models adopt α-helical secondary structure in water + TFE environment. This observation is consistent with both laboratory circular dichroism (CD) experiments and previous molecular dynamics simulation results for alt-SNAPP[13]. Notably, the lipidated SNAPP models exhibit a higher α-helical content in water + TFE environment compared to their non-lipidated counterparts. The increased helical content in the lipidated models suggests that the hydrophobic interactions introduced by the lipid chains contribute to a more stable secondary structure.

These observations are also consistent with findings in the literature, which demonstrate that lipidation, namely, attaching lipid moieties to antimicrobial peptides (AMPs), significantly enhances their α-helical structure. This structural augmentation is critical for their interaction with and disruption of microbial membranes, improving their overall antimicrobial efficacy. For instance, Chu-Kung et .al[51] demonstrated that fatty acid conjugation promotes the antimicrobial activity of peptides by enhancing their membrane affinity and stability. Also Rounds et .al [52] showed that lipidation directly increases the helical content of AMPs, enhancing their stability and membrane-disrupting properties, making it a promising strategy for future drug design. Similarly, Lockwood et .al[53] reported that acylation of the SC4 dodecapeptide increases its bactericidal potency, particularly against Gram-positive bacteria, including drug-resistant strains. Furthermore, studies have shown that lipidation of naturally occurring α-helical AMPs, such as LL-37, increases their α-helicity and improves their antimicrobial activity[19].

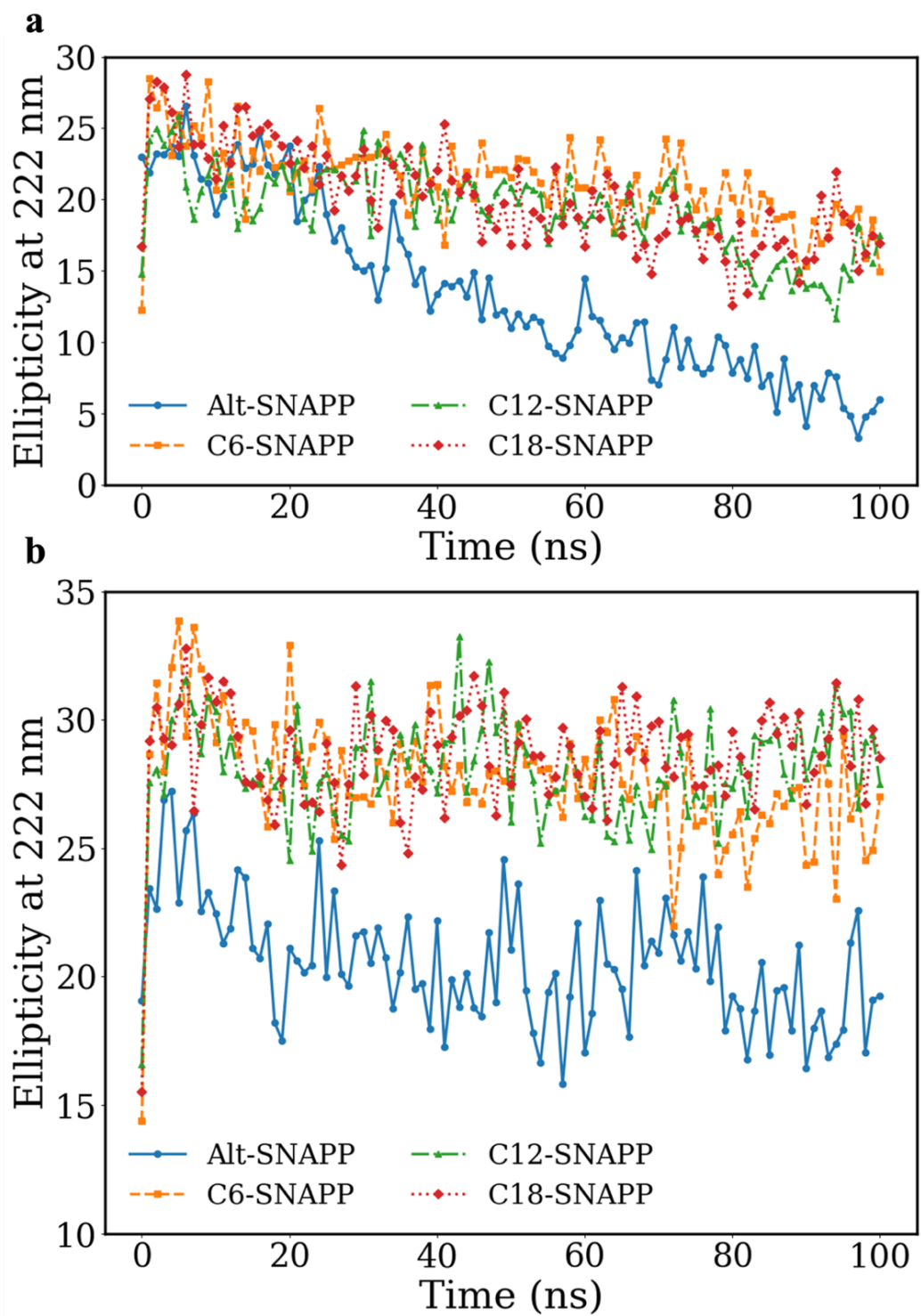

*Figure 3 – a) Ellipticity at 222 nm for alt-SNAPP, C6-SNAPP, C12-SNAPP, and C18-SNAPP in water (averaged over all eight arms), demonstrating that while the helicity of lipidated SNAPP models remains higher than that of alt-SNAPP, the helicity decreases over time in the hydrophilic environment (water). b) Ellipticity at 222 nm for alt-SNAPP, C6-SNAPP, C12-SNAPP, and C18-SNAPP in water + TFE, showing that in a hydrophobic environment (water + TFE), the lipidated SNAPP models exhibit higher helical content compared to alt-SNAPP, with helicity remaining stable over time.*

### 3.2. Mechanism of Action

### 3.2.1. C6-SNAPP and C12-SNAPP bind strongly and deform membrane bilayers

As described in the methodology, C6-SNAPP was initially placed 2 nm above the lipid bilayer, and the simulation was run for 1 μs. Figure 4 provides a summary of the analysis performed on this simulation, where subfigures 4a to 4d present snapshots at different time intervals (0, 250, 500, and 1000 ns). Additionally, subfigures 4e and 4f display the contact analysis and bilayer thickness analysis, respectively. As shown in Figure 4a, the initial configuration of C6-SNAPP remains 2 nm above the bilayer surface with no penetration into the bilayer. The molecule interacts with the bilayer primarily through electrostatic attraction, but unlike alt-SNAPP, it retains its star shape without flattening. In our previous study on the alt-SNAPP model[13], we demonstrated that alt-SNAPP interacts with the bilayer through electrostatic attraction between lysine residues and lipid head groups. However, this interaction triggers a structural transition from a star-like shape to a flattened conformation, which subsequently enhances hydrogen bonding with the membrane.

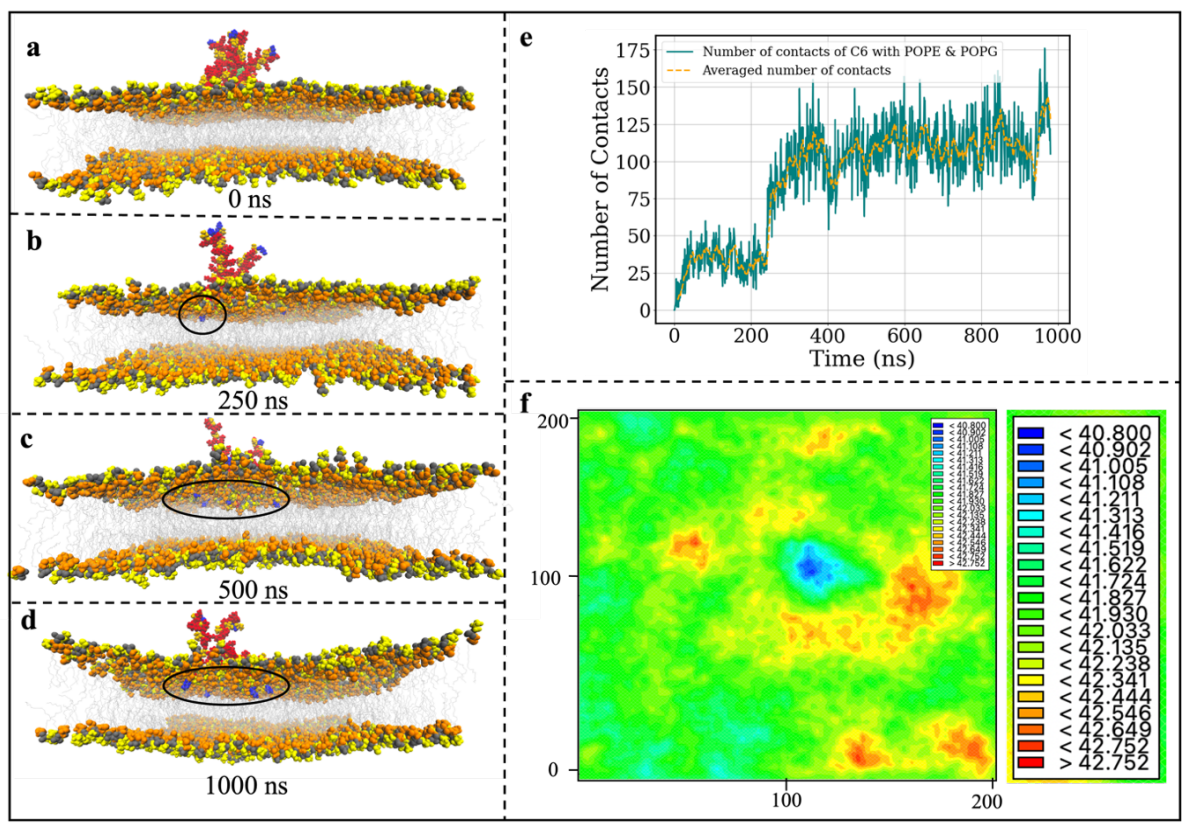

*Figure 4 - Analysis of interaction of model C6-SNAPP with lipid bilayer. The phosphate group (phosphate and oxygen atoms) is coloured in grey, the glycerol backbone (C1, C2, and C3 carbon atoms) is coloured in yellow, and the head group (N and attached hydrogen atoms) is coloured in orange. The lipid tails are shown in transparent silver, and water is excluded from the snapshots for clarity. (a) Snapshot of the initial configuration of C6-SNAPP at 0 ns. (b) Snapshot at 250 ns, showing one C6 group beginning to enter the hydrophobic core. (c) Snapshot at 500 ns, showing three C6 groups starting to penetrate the hydrophobic core. (d) Snapshot at 1000 ns, showing four C6 groups fully inserted into the hydrophobic core of the bilayer. (e) Number of contacts made by C6 groups with POPE and POPG lipids. (f) Membrane thickness map from VMD, illustrating changes in the membrane thickness over time (averaged over the 1000ns), with dark blue regions indicating thinning ( < 40.8 Å), red regions indicating thickening ( > 42.75 Å), and green regions representing unaffected areas of the membrane.*

The initial stages of the simulation of C6-SNAPP reveal that within the first 10 ns, the C6 group of a C6-SNAPP arm closest to the bilayer surface begins to penetrate the membrane. Figure 4e shows that at this early stage, C6 forms approximately 30 contacts with the bilayer, with the number of contacts gradually increasing over time. By 250 ns, as illustrated in Figure 4b, this C6 group has fully embedded itself within the hydrophobic core. Over the course of the simulation, the number of C6 groups of the arms penetrating the hydrophobic core increases progressively, as shown in Figures 4a, 4b, 4c, and 4d, with the C6 portions of these arms fully entering the lipid hydrophobic core. By the end of the simulation, all four arms near to the bilayer have completely penetrated the hydrophobic region, as depicted in Figure 4d. This interaction is further substantiated by Figure 4e, which reveals that C6 forms approximately 130 contacts with the bilayer within 1 μs of simulation. Figure 4f presents a membrane thickness map obtained from VMD membrane analysis tool and highlights the effects of C6-SNAPP on the bilayer's thickness, showing regions of thinning (blue regions) and compression (red regions). These structural alterations demonstrate that C6-SNAPP induces moderate disruptions to membrane thickness, lipid packing, and bilayer integrity.

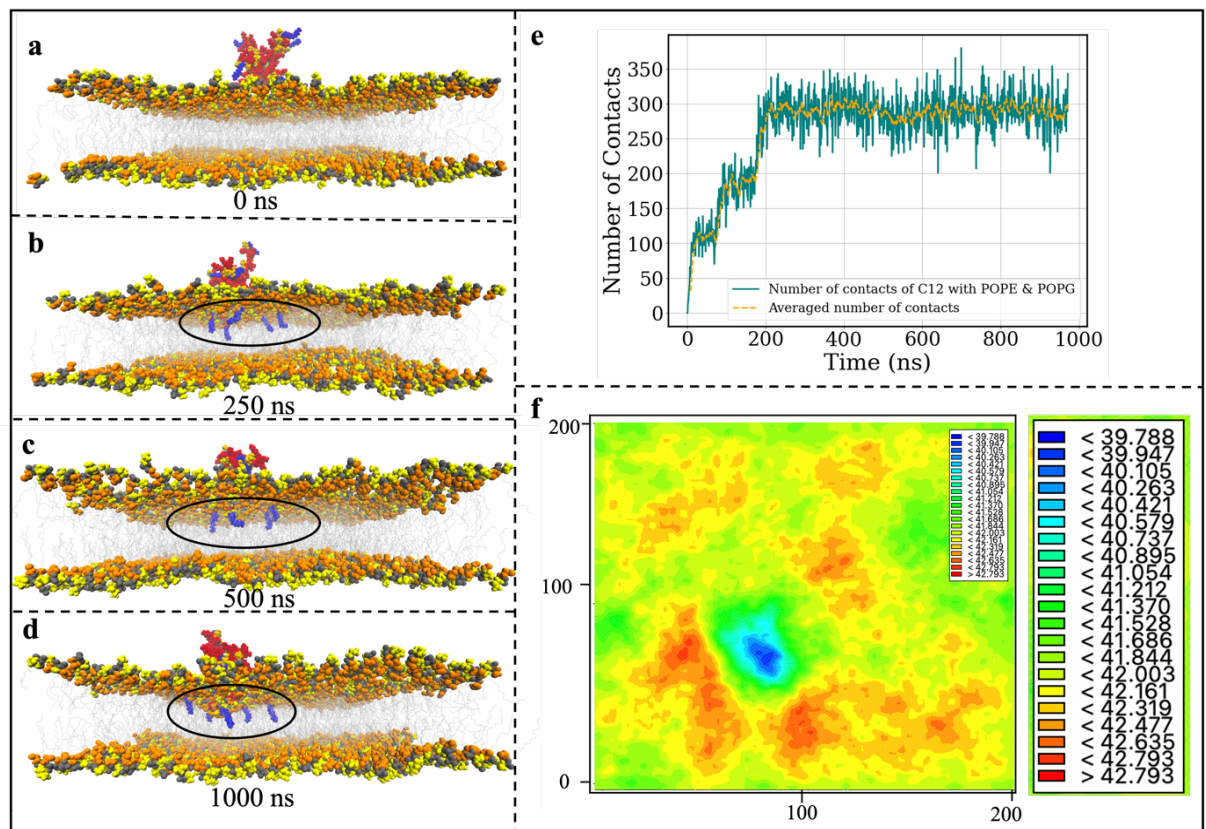

*Figure 5 - Analysis of the interaction of model C12-SNAPP with lipid bilayer. The phosphate group (phosphate and oxygen atoms) is colored in grey, the glycerol backbone (C1, C2, and C3 carbon atoms) in yellow, and the head group (N and attached hydrogen atoms) in orange. The lipid tails are shown in transparent silver, and water is excluded from the snapshots for clarity. (a) Snapshot of the initial configuration of C12-SNAPP at 0 ns. (b) Snapshot at 250 ns, showing four C12 groups entering the hydrophobic core. (c) Snapshot at 500 ns, showing further progression of C12 groups. (d) Snapshot at 1000 ns, illustrating C12 groups fully interacting with the hydrophobic core of the bilayer. (e) Number of contacts made by C12 groups with POPE and POPG lipids. (f) Membrane thickness map from VMD, showing changes in membrane thickness over time, with dark blue regions indicating thinning ( < 39.788 Å), red regions indicating thickening ( > 42.793 Å), and green regions representing unaffected membrane areas.*

Similarly, as shown in Figure 5, C12-SNAPP induces significant bilayer thinning. Initially, as illustrated in Figure 5a, the C12-SNAPP was positioned 2 nm above the bilayer. Over time, as depicted in Figure 5b, all four arms of the C12-SNAPP, including the entire C12 group, became embedded in the hydrophobic core of the bilayer. This behaviour contrasts with the C6-SNAPP, where only one arm penetrated the bilayer within 250 ns. At 500 ns (Figure 5c) and 1000 ns (Figure 5d), the inserted C12 groups exhibited strong interactions with the lipid tails. This trend is quantitatively represented in the contact graph (Figure 5e), which highlights three distinct phases of increasing contacts: from 0 to 100 within the first 100 ns, 100 to 150 between 100 ns and 200 ns, and 150 to 200 over the remaining 200 ns to 1000 ns. The number of contacts for C12-SNAPP is significantly higher than that of C6-SNAPP, with the maximum reaching approximately 300 which is expected as C12 is twice the length of C6.

Additionally, C12-SNAPP induces considerable bilayer thinning, as indicated by the blue regions in the bilayer thickness map, alongside significant bilayer compression (red regions) throughout the system. The extent of bilayer thinning caused by C12-SNAPP is more pronounced compared to C6-SNAPP, as evidenced visually and by the quantitative bilayer thickness values shown in Figures 5f and 4f respectively.

These suggest that lipidation, i.e., the conjugation of lipid chains to peptides enhances antimicrobial peptide (AMP) activity by improving peptide-membrane interactions, facilitating deeper insertion into the lipid bilayer, and stabilizing peptide binding. Lipidation of SNAPPs provides additional functionality by enabling the arms to embed more effectively into the membrane, disrupting the lipid bilayer structure through stress and destabilization[17, 21, 24, 29, 31]. However, the effect of lipid chain length is nuanced; short chains like hexanoic acid (C6) and moderate chains like lauric acid (C12) exhibit distinct behaviours, which influence their interaction with bacterial membranes.

The impact of lipid chain length on AMP activity has been extensively studied. Moderate chain lengths, such as C6–C12, have been shown to enhance membrane interaction and antimicrobial potency[25, 30, 51, 52, 54]. For example, Chu-Kung et al.[51] demonstrated that fatty acid conjugation enhances the antimicrobial activity of peptides by promoting α-helical formation and improves membrane insertion. Rounds et al.[52] showed that lipidation directly increases the helical content of AMPs, enhancing their stability and membrane-disrupting properties, making it a promising strategy for drug design. Makowska et al.[30] found that lipidation significantly enhances the α-helical content of antimicrobial peptides (AMPs), which improves their stability, membrane-binding affinity, and overall antimicrobial activity, making it a promising strategy for developing effective therapeutic agents. Radzishevsky et al.[54] found that acylation with moderate chain lengths improves activity against Gram-positive bacteria but may reduce activity against Gram-negative bacteria.

Moreover, the insertion of lipidated peptides into the membrane can trigger additional bactericidal mechanisms. By inducing mechanical stress and disrupting lipid packing, lipidated peptides can activate programmed cell death pathways, such as autolysis and reactive oxygen species production[26]. Additionally, their ability to destabilize lipid domains within membranes further amplifies their antimicrobial action[24, 55].

Both C6-SNAPP and C12-SNAPP demonstrate the ability to interact with bacterial membranes through the insertion of their lipidated termini. Molecular dynamics simulations reveal that the lipid tails in both variants penetrate the bilipid membrane rapidly, embedding within the hydrophobic core of the bilayer. As shown in Figures 4f, 5f and 7, this insertion induces localized mechanical stress, disrupts lipid packing, and destabilizes the bilayer, and could potentially impair membrane function as discussed in references [26, 55].

## 3.2.2. C18-SNAPP adopts a highly compact structure and weakly deforms membranes

C18-SNAPP features stearic acid (C18 lipid) attached to the termini of its arms. For C18-SNAPP, simulations reveal that longer lipid chains do not immediately insert into the hydrophobic core of the bilipid membrane as shown in Figure 6a and 6b. Instead, arms of the molecule back-folds inward, forming a compact, blob-like structure, as shown in Figure 6g. This back-folding limits effective interaction with the lipid bilayer and hence hinders its ability to disrupt bacterial membranes. After approximately 300 ns, one C18 group begins to penetrate the bilayer's hydrophobic core, as visually depicted in Figure 6c. This behaviour is also evident in the contact analysis, which shows a significant increase in the number of contacts between 300 ns and 500 ns.

As illustrated in Figure 6d, the C18 group of C18-SNAPP bends excessively while interacting with the lipid tails, resulting in a decline in the number of contacts toward the end of the simulation. Furthermore, the bilayer thickness map indicates (Figure 6f) no significant bilayer thinning caused by C18-SNAPP. The rest of the contact map appears predominantly green, suggesting minimal bilayer compression across the membrane. This behaviour contrasts with the C6-SNAPP and C12-SNAPP models, where notable bilayer thinning and compression were observed. Among the three SNAPP models, C18-SNAPP exhibits the least bilayer thinning.

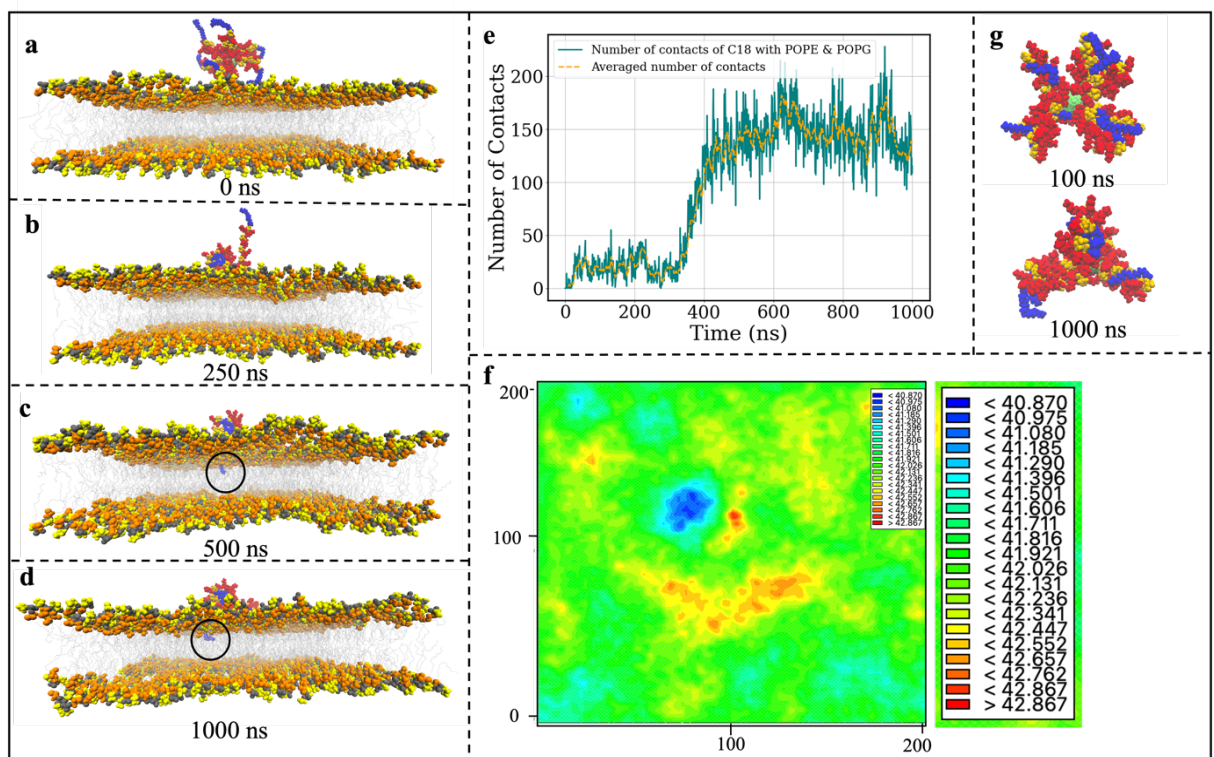

*Figure 6 - Analysis of the interaction of the model C18-SNAPP with lipid bilayer. The phosphate group (phosphate and oxygen atoms) is coloured in grey, the glycerol backbone (C1, C2, and C3 carbon atoms) in yellow, and the head group (N and attached hydrogen atoms) in orange. The lipid tails are shown in transparent silver, and water is excluded from the snapshots for clarity. (a) Snapshot of the initial configuration of C18-SNAPP at 0 ns. (b) Snapshot at 250 ns, showing no C18 groups entering the hydrophobic core. (c) Snapshot at 500 ns, showing one C18 group starting to enter the hydrophobic core. (d) Snapshot at 1000 ns, showing the entered C18 group bending as it interacts with the bilayer tails. (e) Number of contacts made by C18 groups with POPE and POPG lipids. (f) Membrane thickness map from VMD, showing changes in membrane thickness over time, with dark blue regions indicating thinning (<40.870 Å), red regions indicating thickening (>42.867 Å), and green regions representing unaffected membrane areas.*

The inward bending of the lipidated arms in C18-SNAPP is likely a result of hydrophobic interactions between the stearic acid chains themselves. Such self-interaction is consistent with findings from other studies on lipidated peptides, where overly long lipid chains promote intra-peptide interactions rather than membrane engagement[25, 56]. This phenomenon reduces the availability of the lipid termini for interaction with the bilayer and disrupts the multivalent engagement typically observed in SNAPPs. While longer lipid chains generally increase hydrophobicity and membrane affinity[51, 56], excessive hydrophobicity can lead to undesirable effects such as self-association of SNAPP arms through inward bending, which may reduce effective membrane interaction.

Figure 7 presents the upper leaflet deformation profiles for C6-SNAPP, C12-SNAPP, and C18-SNAPP, illustrating the extent of membrane thinning and thickening induced by each peptide. The results indicate that the most significant deformation occurs with C12-SNAPP, where the upper leaflet undergoes thinning of less than -4.72 Å and thickening exceeding 1.56 Å. In contrast, C6-SNAPP exhibits a lower degree of deformation, with thinning of less than -3.32 Å and thickening greater than 1.37 Å. As previously noted, C18-SNAPP induces the least deformation, with thinning restricted to less than 1.02 Å and thickening exceeding only 1.02 Å.

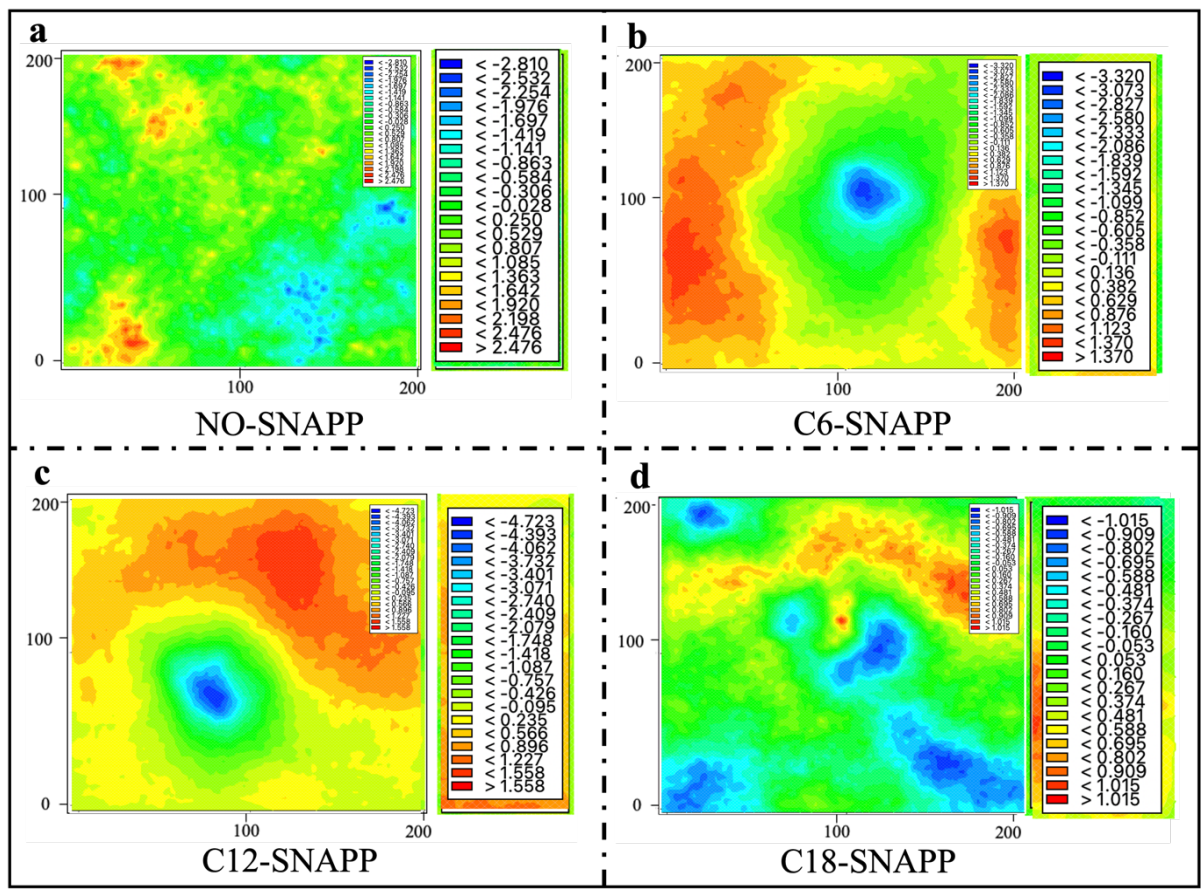

*Figure 7 - Upper leaflet deformation of the POPE/POPG bilayer as analysed using the VMD Membrane Analysis Tool. The deformation map illustrates upper leaflet deformations over time, where dark blue regions indicate membrane thinning, red regions indicate thickening, and green regions represent unaffected membrane areas. (a) Displays the bilayer fluctuations without any SNAPP model (b) Displays the highest upper leaflet deformation due to C6-SNAPP interaction with the bilayer. (c) Represents moderate upper leaflet deformation caused by C12-SNAPP interaction. (d) Depicts the lowest upper leaflet deformation resulting from C18-SNAPP interaction with the bilayer.*

## 3.3. Steered Molecular Dynamics and Potential of Mean force Analysis

### 3.3.1. C6 & C12 lipidation reduces resistance to SNAPP membrane permeation

Pulling force is a measure of resistance encountered when a molecule or a molecular fragment, such as one arm of SNAPP is forcibly moved along a specific trajectory, often through a lipid bilayer. This is achieved by applying a harmonic bias potential along a collective variable, often the reaction coordinate, which represents the displacement of the molecule. The pulling force is a direct function of the energy required to overcome molecular interactions, such as hydrophobic, electrostatic, and hydrogen bonding between the molecule and its environment. By analysing the force profile, one can gain critical insights into the interactions governing the mechanism of action of SNAPP molecules and its interaction with the lipid bilayer.

The pulling force profiles for alt-SNAPP, C6-SNAPP, C12-SNAPP and C18-SNAPP arms displayed in Figure 8 provide valuable insights into their interactions with the lipid bilayer and their respective mechanisms of action. The x-axis represents the pulling distance across the membrane, while the y-axis denotes the force in kJ/mol/nm. The force remains low in the aqueous regions (-4 to -3 nm and beyond 3 nm) as both SNAPP arms experience minimal resistance in the water phase. As the molecules enter the lipid headgroup regions (-3 to -2 nm and 2 to 3 nm), the force starts to rise due to electrostatic and hydrogen bonding interactions between the SNAPP arms and the hydrophilic head groups.

The alt-SNAPP arm (Figure 8, blue line) exhibits the highest force peak of 884.53 kJ/mol/nm at x of 2.25 nm, indicating that it experiences the strongest resistance within the lower leaflet head group region of the bilayer. The force then decreases as the peptide exits into the aqueous phase of the lower leaflet. In comparison, the C6-SNAPP arm (Figure 8, green line) reaches a lower maximum force of 855.00 kJ/mol/nm at x of 0.10 nm, (within the bilayer core) suggesting that the addition of a short lipid chain slightly decreases the force required to pull the peptide into the bilayer.

For C12-SNAPP (Figure 8, orange line), the force peak is further reduced to 815.17 kJ/mol/nm at x of 1.40 nm, reinforcing the trend that increasing hydrophobicity decreases the required force to pull the peptide into the bilayer. The peak shift indicates that C12-SNAPP experiences its greatest resistance within the lower leaflet of the bilayer but at a different depth compared to alt-SNAPP and C6-SNAPP. The lower force requirement suggests that the longer lipid tail enhances hydrophobic interactions, reducing the energy barrier for translocation.

The C18-SNAPP arm (Figure 8, magenta line) presents a distinct profile, with a force peak of 883.11 kJ/mol/nm at x of 3.24 nm, and a broader distribution compared to the other SNAPP variants. This broad peak suggests prolonged interactions with the bilayer. While lipidation enhances membrane permeability by facilitating bilayer traversal, excessive hydrophobic modification may lead to prolonged bilayer association, which could influence peptide-membrane interactions in biological applications. These results are consistent with findings that lipidated peptides enhance membrane disruption through improved compatibility with the lipid bilayer[25, 57, 58].

Figure 9 offers a clear visual representation of how different lipidated SNAPP arms interact with and deform the lipid bilayer during steered molecular dynamics simulations. When paired with the force profile data, these structural snapshots give valuable insight into how lipid tail length and composition influence bilayer disruption.

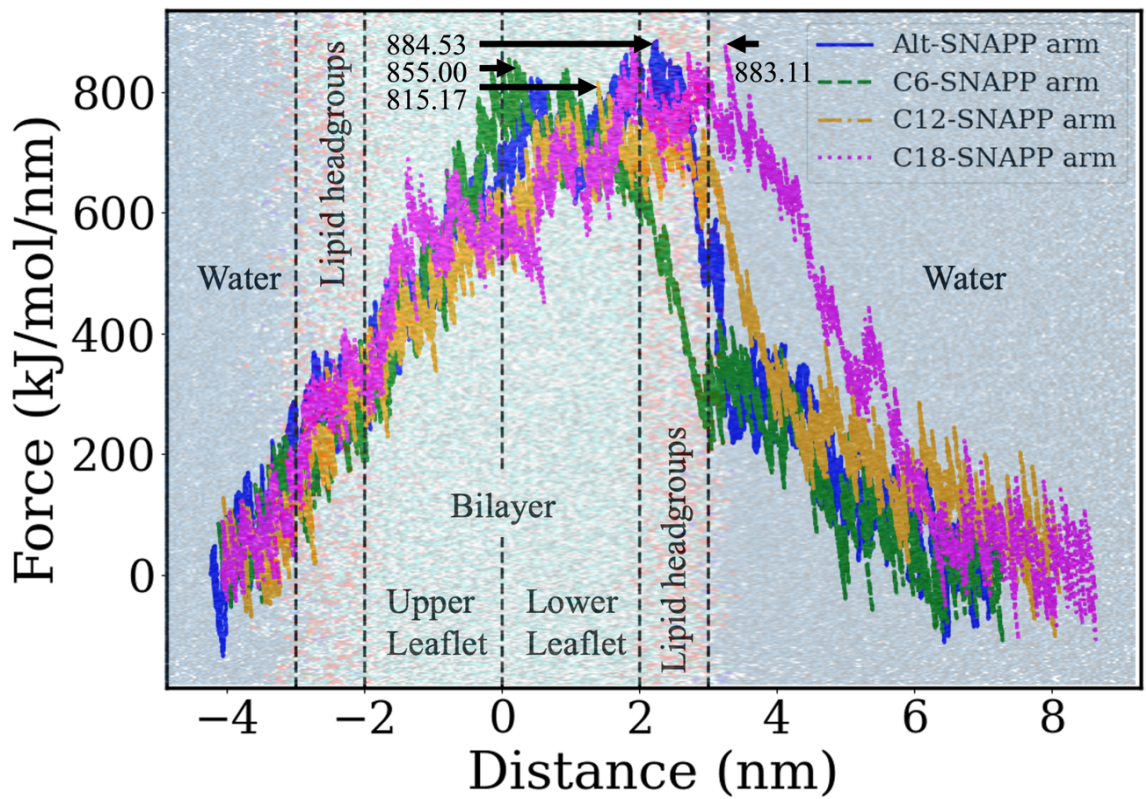

*Figure 8 - Pulling force profiles for alt-SNAPP arm, C6-SNAPP, C12-SNAPP and C18-SNAPP arms. Detailed methodologies for the pulling simulations are described in the methods section.*

Among all variants, the alt-SNAPP arm as shown in Figure 9, appears to induce the most dramatic bilayer deformation. As it pulls through the membrane, it drags lipid headgroups from the upper leaflet deep into the lower leaflet, creating a noticeable V-shaped indentation (Figure 9c). This correlates well with its high force peak of 884.53 kJ/mol/nm at 2.25 nm, suggesting that the peptide encounters strong resistance as it moves through the densely packed head group region of the lower leaflet. This kind of deformation could reflect potent membrane-disruptive behaviour, tied to its antimicrobial activity, which we reported in our previous study[13].

The C6-SNAPP, in comparison, shows much milder effects on the bilayer. The structural deformation is less pronounced as shown in Figure 9, and the force peak is slightly lower at 855.00 kJ/mol/nm, occurring near the bilayer centre (0.10 nm). This might suggest that adding a short lipid tail reduces the energy barrier for insertion without strongly perturbing the membrane structure. Interestingly, the C12-SNAPP arm follows the same trend: minimal structural disruption and a further drop in the force peak to 815.17 kJ/mol/nm, peaking at 1.40 nm.

The C18-SNAPP arm, however, presents a more complex picture. Visually as shown in Figure 9, it causes substantial bilayer disruption, and its force peak 883.11 kJ/mol/nm at 3.24 nm is high but broader compared to the others. This broader distribution could indicate prolonged interactions or delayed disengagement from the bilayer.

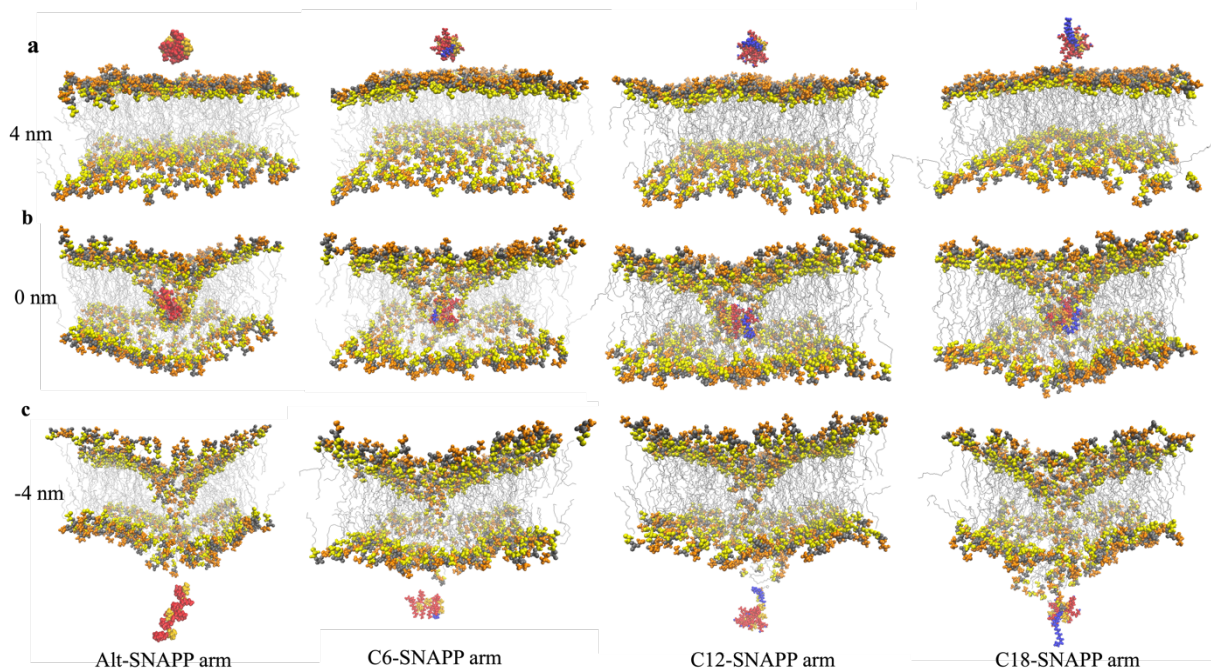

*Figure 9 – Snapshots of bilayer deformations observed during steered molecular dynamics simulations. Panels a, b, and c correspond to positions 4 nm above the bilayer centre, at the centre (0 nm), and 4 nm below the centre, respectively. Results are shown for all four scenarios: alt-SNAPP arm, C6-SNAPP arm, C12-SNAPP arm, and C18-SNAPP arm. The phosphate group (phosphate and oxygen atoms) is coloured grey, the glycerol backbone (C1, C2, and C3 carbon atoms) is yellow, and the head group (nitrogen and attached hydrogen atoms) is orange. The bilayer lipid tails are shown in transparent silver, lysine residues are shown in red, valine residues in yellow, lipidated ends are in blue and water is excluded from the snapshots for clarity.*

### 3.3.2. Potential of mean force profiles reveal lipidation-dependent reduction in permeation free energy

The potential of mean force (PMF) profiles displayed in Figure 10 for the translocation of the alt-SNAPP arm, C6-SNAPP arm, C12-SNAPP arm and C18-SNAPP arm through a POPE/POPG bilayer reveal critical differences in their membrane interaction and penetration behaviours. All the peptides show an initial free energy stabilization at approximately -3 nm, indicative of adsorption onto the bilayer interface. This stabilization is driven by electrostatic interactions between the positively charged residues of the peptides and the negatively charged lipid headgroups of POPE and POPG, consistent with previously observed trends in antimicrobial peptide-membrane interactions[59, 60].

The key distinction between the SNAPP arm models is observed in the energy barrier associated with their translocation through the hydrophobic bilayer core. As shown in Figure 10, in the initial stage at point "a", the alt-SNAPP arm is in the bulk water phase with a potential of mean force (PMF) value of 0 kJ/mol. As the peptide moves closer to the bilayer interface, it encounters interactions with the lipid headgroups, leading to a local energy minimum at "b" (Figure 10), where the energy drops to -28.9 kJ/mol. This negative energy suggests a favourable adsorption of the alt-SNAPP arm onto the bilayer surface, primarily driven by electrostatic and van der Waals interactions between the positively charged lysine residues of the peptide and the negatively charged phosphate groups in the lipid headgroups. Additionally, hydrophobic interactions between the peptide's nonpolar regions and the lipid tails further stabilize the adsorption consistent with previously observed trends in antimicrobial peptide-membrane interactions[59, 61]. Similar trend is shown for lipidated SNAPP arms as well. This phase is crucial as it sets up the initial interaction, which determines the peptide's subsequent penetration into the bilayer.

As the alt-SNAPP arm transitions from "b": (-28.9 kJ/mol) to "c" (381 kJ/mol), the energy drastically increases by 409.9 kJ/mol, indicating a significant energy barrier that must be overcome to translocate through the hydrophobic core of the bilayer as shown in the Figure 10. This steep rise in energy results from several factors. First, mutual electrostatic repulsion occurs between the positively charged lysine residues due to the lack of hydration as they pass through the bilayer's hydrophobic interior. Second, hydrophobic core resistance presents a substantial challenge, as the nonpolar lipid tails strongly oppose the movement of polar or charged groups, leading to an energy penalty as the peptide traverses this region. Third, lipid rearrangement occurs as the insertion of the alt-SNAPP arm distorts the lipid packing, requiring additional energy to overcome bilayer deformation. Finally, hydration shell disruption adds to the free energy cost, as water molecules that typically stabilize charged residues must be displaced, making the passage through the membrane even more energetically demanding. This energy peak at "c" (381 kJ/mol) represents the highest resistance point for the peptide's translocation. From "c" (381 kJ/mol) to "d" (345 kJ/mol), there is a decrease of 36 kJ/mol as the peptide moves past the second lipid headgroup layer and into the water phase as shown in the Figure 10. This decline in energy can be attributed to several factors. First, rehydration of lysine residues occurs as the peptide exits the bilayer, allowing its charged groups to regain hydration, which stabilizes the structure and lowers the energy requirement. Second, membrane reorganization takes place as the alt-SNAPP arm fully translocates, allowing the surrounding lipids to return to a lower-energy conformation. Third, reduced hydrophobic resistance plays a role, as the peptide is no longer embedded in the nonpolar region, leading to stabilization.

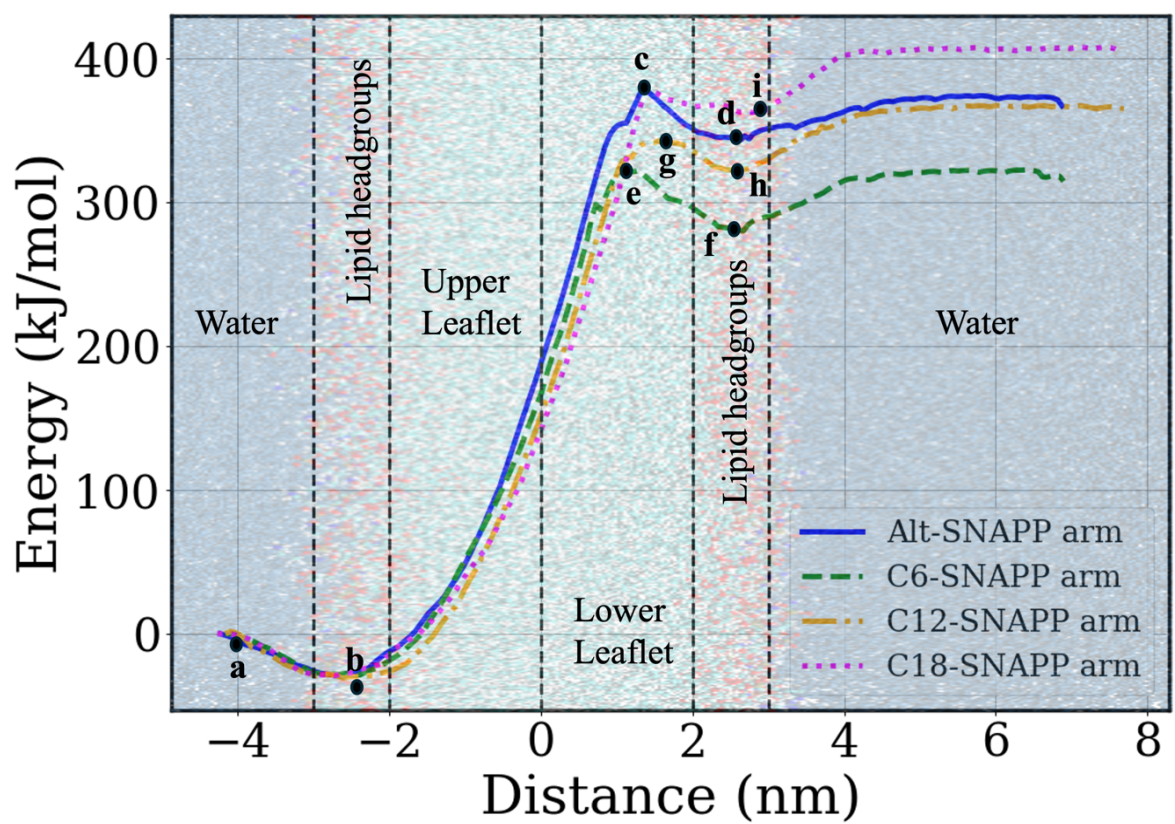

*Figure 10 - Potential mean force (PMF) calculated for the alt-SNAPP, C6-SNAPP, C12-SNAPP and C18-SNAPP arms using the umbrella sampling method with weighted histogram analysis, demonstrating that the PMFs for the C6-SNAPP arm and C12 SNAPP arm are lower than that of the alt-SNAPP arm.*

In contrast, the C6-lipidated SNAPP arm exhibits a reduced energy barrier of 350.9 kJ/mol, emphasizing the role of lipidation in facilitating peptide penetration through the bilayer, as observed in Figure 10, points "b" to "e". Similarly, the C12-SNAPP arm experiences an energy barrier of 362.9 kJ/mol along the translocation pathway from points "b" to "g", reinforcing the trend that moderate lipidation enhances bilayer compatibility and reduces the energetic cost of membrane insertion. This observation supports the hypothesis that hydrophobic modifications improve compatibility with the bilayer core, as previously reported in the literature[62]

Notably, akin to alt-SNAPP, both C6 and C12-lipidated SNAPP arms exhibit a significant energy reduction after reaching the lower leaflet lipid headgroup, with energy decreasing by 42 kJ/mol (e to f) for C6-SNAPP and 11 kJ/mol ("g" to "h") for C12-SNAPP.

However, in contrast to C6 and C12-SNAPP, the C18-SNAPP variant exhibits an energy barrier comparable to that of alt-SNAPP, indicating that excessive lipidation does not further reduce the energy cost of bilayer penetration. The energy decreases upon reaching the lower leaflet lipid headgroup to 18 kJ/mol, suggesting that while C18-SNAPP maintains strong bilayer interactions, it does not exhibit the same level of facilitated translocation observed in C6 and C12-SNAPP variants.

The lower energy barrier for C6 and C12-SNAPP indicates enhanced translocation efficiency and an increased likelihood of achieving membrane disruption. This property is particularly relevant in the context of antimicrobial activity, as faster and more energetically favourable peptide insertion correlates with increased efficacy.

## 4. Conclusion

This study provides molecular insights into the structural and functional implications of lipidation on Structurally Nano Engineered Antimicrobial Peptide Polymers. By examining the secondary structure, mechanism of action, pulling force profiles, and potential of mean force for alt-SNAPP and lipidated variants (C6-SNAPP, C12-SNAPP, and C18-SNAPP), we elucidate the profound effects of lipid modifications on the antimicrobial activity of these peptides.

The secondary structure analysis reveals that lipidation enhances the α-helical stability of SNAPP arms in both hydrophilic (water) and in hydrophobic (water + TFE) environments. This stability is particularly evident in C6-SNAPP and C12-SNAPP, where lipid modifications promote hydrophobic interactions that preserve the α-helical conformation, a key determinant of membrane interaction and disruption. Conversely, the longer lipid chains in C18-SNAPP induce back-folding of arms, limiting its ability to effectively interact with bacterial membranes. As demonstrated by the bilayer thickness maps and upper leaflet deformation maps, C12-SNAPP causes the greatest disruption, followed by C6-SNAPP, with C18-SNAPP exhibiting the least effect on the bilayer.

Mechanistic insights from the pulling force simulations highlight the contrasting behaviours of alt-SNAPP, C6-SNAPP, C12-SNAPP and C18-SNAPP arms. Alt-SNAPP arm and C18-SNAPP arm exhibit slightly higher pulling force peaks, indicating resistance during translocation across the bilayer, driven by its balanced hydrophilic and hydrophobic interactions. In contrast, C6-SNAPP and C12-SNAPP demonstrate a reduced pulling force, underscoring the role of lipidation in facilitating deeper insertion into the membrane. This enhanced compatibility with the bilayer's hydrophobic core disrupts lipid packing and induces localized mechanical stress, contributing to its bactericidal activity.

The PMF profiles further emphasize the impact of lipidation on membrane penetration. The regular alt-SNAPP and the longer lipidated C18-SNAPP arms encounter a higher energy barrier of 409.9 kJ/mol during translocation, reflecting strong resistance from the hydrophobic bilayer core. The lipidated C6-SNAPP and C12-SNAPP, however, show a lower energy barriers of 350.9 kJ/mol ad 362.9 kJ/mol respectively, highlighting the effectiveness of lipidation in improving bilayer compatibility. This reduction in energy cost not only enhances the peptide's ability to disrupt bacterial membranes but also accelerates the translocation process, increasing its antimicrobial potency.

Overall, this study underscores the critical role of lipidation in optimizing the structure and function of SNAPPs. By reducing the energy barrier for translocation, lipidation enhances membrane activity, providing a foundation for further designing next-generation antimicrobial peptides. The findings highlight the importance of tuning lipid chain length to balance membrane interaction and avoid back-folding of arms and collapsing, offering valuable design principles for combating multidrug-resistant bacterial strains. Future work should explore alternative lipid modifications and their effects on peptide efficacy across diverse bacterial species using advanced machine learning algorithms[63,64,65,66] and high-throughput MD simulations[67].

## Acknowledgment

A.J is supported by research training programme (RTP) scholarship provided by the Australian government. This research was supported by The University of Melbourne's Research Computing Services and the Petascale Campus Initiative.

## Conflict of interest

Authors declare no conflict of interest.

## Appendix

Table A1 – Molecular Dynamics simulation details for secondary structure analysis, bilayer simulations, and umbrella sampling analysis.

|  | Box size (nm) | Number of water molecules (TIP3) | Number of TFE or POPE & POPG molecules | Number of ions | Simulation time (ns) |
| --- | --- | --- | --- | --- | --- |
| **Secondary structure analysis** | | | | | |
| **Alt-SNAPP in water** | 7.75341<br>7.75341<br>7.75341 | 14665 | - | 80 | 100 |
| **Alt-SNAPP in water + TFE** | 7.75341<br>7.75341<br>7.75341 | 2666 | TFE - 2650 | 80 | 100 |
| **C6-SNAPP in water** | 8.93330<br>8.93330<br>8.93330 | 22199 | - | CLA - 80 | 100 |
| **C6-SNAPP in water + TFE** | 8.93330<br>8.93330<br>8.93330 | 4317 | TFE - 3700 | CLA - 80 | 100 |
| **C12-SNAPP in water** | 11.13210<br>11.13210<br>11.13210 | 43955 | - | CLA - 80 | 100 |
| **C12-SNAPP water + TFE** | 11.13210<br>11.13210<br>11.13210 | 7741 | TFE - 7400 | CLA - 80 | 100 |
| **C18-SNAPP in water** | 11.53190<br>11.53190<br>11.53190 | 49344 | - | CLA - 80 | 100 |
| **C18-SNAPP in water + TFE** | 11.53190<br>11.53190<br>11.53190 | 9188 | TFE - 8200 | CLA - 80 | 100 |
| **Bilayer simulations** | | | | | |
| **C6-SNAPP in POPE and POPG bilayer** | 20.03050<br>20.03050<br>18.00000 | 178462 | POPE – 1080<br>POPG – 270 | POT - 755<br>CLA - 565 | 1000 |
| **C12-SNAPP in POPE and POPG bilayer** | 20.03050<br>20.03050<br>18.00000 | 178454 | POPE – 1080<br>POPG - 270 | POT – 755<br>CLA - 565 | 1000 |
| **C18-SNAPP in POPE and POPG bilayer** | 20.03050<br>20.03050<br>18.00000 | 178386 | POPE – 1080<br>POPG - 270 | POT – 755<br>CLA - 565 | 1000 |
| **Umbrella sampling simulations** | | | | | |
| **Alt-SNAPP arm** | 10.05226<br>10.05226<br>14.80648 | 33766 | POPE – 272<br>POPG - 68 | POT – 149<br>CLA - 91 | - |
| **C6-SNAPP arm** | 10.05226<br>10.05226<br>16.91381 | 40958 | POPE – 272<br>POPG - 68 | POT – 149<br>CLA - 91 | - |
| **C12-SNAPP arm** | 10.05226<br>10.05226<br>16.81299 | 40442 | POPE – 272<br>POPG - 68 | POT – 167<br>CLA - 109 | - |
| **C18-SNAPP arm** | 10.05226<br>10.05226<br>16.96830 | 40940 | POPE – 272<br>POPG - 68 | POT – 169<br>CLA - 111 | - |